\documentclass{gistt}

\usepackage{paralist}
\usepackage{listings}
\usepackage{todonotes}
\usepackage{xspace}

\usetikzlibrary{calc}

\bibliography{quellen.bib}

\title{Overhead Measurement Noise in Different Runtime Environments}
\author{David Georg Reichelt\\
Lancaster University Leipzig / \\
Universität Leipzig
\and
Reiner Jung\\
Christian-Albrechts-\\
Universität zu Kiel\\
\and
Andr\'e van Hoorn\\
Universität Hamburg\\
}

\begin{document}

\maketitle

\begin{abstract}
In order to detect performance changes, measurements are performed with the same execution environment. In cloud environments, the noise from different processes running on the same cluster nodes might change measurement results and thereby make performance changes hard to measure.

The benchmark MooBench determines the overhead of different observability tools and is executed continuously. In this study, we compare the suitability of different execution environments to benchmark the observability overhead using MooBench. To do so, we compare the execution times and standard deviation of MooBench in a cloud execution environment to three bare-metal execution environments. We find that bare metal servers have lower runtime and standard deviation for multi-threaded MooBench execution. Nevertheless, we see that performance changes up to 4.41\,\% are detectable by GitHub actions, as long as only sequential workloads are examined.
\end{abstract}

\section{Introduction}

Benchmarking results aim to make performance changes detectable. Therefore, benchmarks are usually executed on the same execution environment, e.g., for application benchmarking, usually the same hardware, operating system, and runtime environment are used. The setup and operation of these execution environments are time-consuming and costly. Due to the availability of cloud execution environments, the time effort for setup and the costs can be reduced. However, cloud environments do not create the same reproducibility of measurements due to the parallel execution of other tasks on the same runtime environment. In this study, we examine which performance changes can be measured both with classical execution environments and with cloud environments.

The MooBench microbenchmark measures the performance overhead of observability frameworks with their different configurations \cite{waller2013benchmark}.\footnote{Formerly, this was called \textit{monitoring overhead}. Due to the widespread use of \textit{observability} in industry, we chose to switched to this term.} MooBench measures the overhead created by an observability tool by executing recursive calls to a method itself, so the only significant resource consumption is caused by the observability tool.
For this study, we exemplarily execute the MooBench microbenchmark on different execution environments: The GitHub actions runner and three different bare metal configurations. We find that
\begin{inparaenum}[(1)]
  \item Measurement durations on GitHub Actions are in the same order of magnitude as modern desktop machines,
  \item The cloud environment is able to detect performance changes of at minimum 4.41\,\%, whereas the bare metal servers are able to detect performance changes of at minimum 1.94\,\% for Kieker's default configuration,
  \item The performance of GitHub Actions is significantly worse for parallel execution, leading to unusability for benchmarking the obversability overhead of parallel threads.
\end{inparaenum}

The remainder of this paper is organized as follows: First, we describe the setup of our measurement, including the GitHub Actions workflow implementation of MooBench. Based on these setups, we describe our measurement results. Subsequently, we compare our study with related work. Finally, we give a summary.% of the paper.

%\section{MooBench}
%MooBench is a benchmark for the overhead of application performance benchmarking \cite{waller2013benchmark}. To azure statistic rigor, MooBench executes a given number of loop starts. In each loop, MooBench sequentially starts a demo application using every monitoring configuration, including no monitoring at all as baseline, and measures the duration of the demo applications execution.

\section{Measurement Setups}

We used four different measurement setups for MooBench: The regular Kieker Jenkins bare metal server, the GitHub Actions workflow, a typical desktop configuration and a Raspberry Pi. For all setups, we limitted the heap space to 2 GB.

Prior measurements have typically been executed on the oldest long-term release version of the JVM, either the JVM 1.7 \cite{knoche2018using}\footnote{Documented in the Zenodo dataset: \url{https://doi.org/10.5281/zenodo.165513}} or JVM 1.8 \cite{reichelt2021overhead} \cite{reichelt2023more}. To benefit from latest performance improvements, we switched to the latest available LTS JVM (JVM 21). This lead to a significant overhead decrease on some platforms (for example on i7-4770 from $3.35 \mu s$ (relative standard deviation $\sigma=2.01\,\%$) to $3.062 \mu s$ ($\sigma=1.55\,\%$).

All measurement setups used the newest available Eclipse Temurin JDK / OpenJDK 21 distribution available on each platform. In the following, we describe the setups of each measurement.
Even though our prior measurement results suggest that the ByteBuddy, DiSL, or Javassist instrumentation creates lower performance overhead \cite{reichelt2024overhead}, we kept MooBenchs default configuration using AspectJ for this study.

\subsection{Kieker Jenkins}

The Kieker Jenkins server uses a slave for performance benchmarking. It is running on Intel Xeon CPU E5620 @ 2.40 GHz (4 cores, 8 threads). The system was running Debian 12 and Eclipse Temurin OpenJDK 21.0.3.

\subsection{Desktop PCs}

We used two Desktop PCs: An i7-4770 @ 3.40GHz (4 cores, 8~threads) running Ubuntu 22.04 and Eclipse Temurin OpenJDK 21.0.2, which we were running before \cite{reichelt2021overhead} \cite{reichelt2023more}, and an Ryzen 5700G @ 3.80 GHz (8 cores, 16~threads), running Rocky Linux 9.4 and Eclipse Temurin OpenJDK 21.0.2.

\subsection{GitHub Actions Workflow}

To automate the performance measurement data management, we use the \textit{github-action-benchmark}\footnote{\url{https://github.com/benchmark-action/github-action-benchmark}}. First, the workflow executes the regular MooBench measurement. Afterwards, the results are transformed into the github-action-benchmark format. Subsequently, the github-action-benchmark action inserts the new measurement run into the measurements and uploads the results to the \lstinline'gh-pages' branch. Finally, the results can be viewed.\footnote{\url{https://kieker-monitoring.github.io/moobench/dev/bench/}}
We used the GitHub Actions Linux runners (4 processes), which are executed by default on a 16 GB memory runner with Ubuntu 22.04. These runners are in the Azure cloud.\footnote{\url{https://docs.github.com/de/actions/using-github-hosted-runners/about-github-hosted-runners/about-github-hosted-runners}} They are running Eclipse Temurin 21.0.3+9.

\subsection{Raspberry Pi}

For comparability with old Raspberry Pi results, we ran the measurements on an Raspberry Pi 4 (4 cores, 4 threads) with 4 GB of RAM. The system was running Raspbian 12 and Eclipse Temurin OpenJDK 21.0.3.

\section{Measurement Results}

In the following, we look at the measurements for MooBench's default parameter and for parallel execution. Our measurement results are available.\footnote{\url{https://zenodo.org/doi/10.5281/zenodo.11355256}}

\subsection{Default Parameter Execution}

Our measurement results are depicted in Table~\ref{tab:results}. We assume that both type I ($\alpha$) and type II ($\beta$) errors should be below 1\,\%. In the minimal change column, we calculated the minimal detectable effect size $\Delta$ of a change with MooBenchs default configuration of $n=10$ loop starts.\footnote{The effect size $\delta$ is the relative change $\Delta$ divided by the relative standard deviation $\sigma$, which can be calculated by  $\delta=\frac{\Delta}{\sigma}=\frac{\sqrt{\frac{n}{2}}-z_{1-\frac{\alpha}{2}}}{z_{1-\beta}}$ \cite{reichelt2022qrs} Therefore, we follow $\Delta = \frac{\sqrt{\frac{n}{2}}-z_{1-\frac{\alpha}{2}}}{z_{1-\beta}} \cdot \sigma$}
While other authors stated that cloud environments are inherently noisy for performance measurement \cite{bulej2020duet, laaber2019software}, we see that for MooBench, GitHub Actions creates measurements with reasonably low noise and at the same time a low execution time. GitHub Actions is faster than the Raspberry Pi and MooBench's Jenkins measurement environment, and comparable to the i7-4770.

\begin{table}[h]
 \setlength{\tabcolsep}{4pt}
 \begin{tabular}{|c|r|r|r|} \hline
             &                            & \multicolumn{1}{c|}{Rel.\,Std.} & \multicolumn{1}{c|}{Min.\,Rel.} \\
  Environment & \multicolumn{1}{c|}{Mean} & \multicolumn{1}{c|}{Deviation $\sigma$}  & \multicolumn{1}{c|}{Change $\Delta$} \\ \hline
  \textbf{Jenkins}    & & & \\
  Baseline      &  108.40 & 4.15\% & 9.28\% \\
%  No Logging    & 2727.28 & 1.76\% & 3.94\% \\
  Binary Writer & 4750.11 & 2.52\% & 5.63\% \\ \hline
  \textbf{GH Actions} & & & \\ 
  Baseline      &   90.68 & 0.11\% & 0.24\% \\
%  No Logging    & 1839.85 & 0.32\% & 0.71\% \\ 
  Binary Writer & 3079.49 & 1.97\% & 4.41\% \\ \hline
  \textbf{Ryzen\,7\,5700G} & & & \\
  Baseline      & 63.84 & 0.16\% & 0.35\% \\
%  No Logging    & 1215.76 & 0.37\% & 0.82\% \\ 
  Binary Writer & 2743.50 & 0.87\% & 1.94\% \\ \hline
  \textbf{i7-4770} & & &  \\
  Baseline      &   55.06 & 0.77\% & 1.73\% \\
%  No Logging    & 1361.81 & 0.35\% & 0.78\% \\
  Binary Writer & 3062.52 & 1.55\% & 3.47\% \\ \hline
  \textbf{Raspberry\,Pi\,4} & & &  \\ 
  Baseline      &   169.64 & 1.39\% & 3.11\% \\
%  No Logging    &  5688.44 & 0.34\% & 0.76\% \\
  Binary Writer & 12595.40 & 4.11\% & 9.20\% \\ \hline
 \end{tabular}
 \caption{Measurement Results (in ns)}
 \label{tab:results}
\end{table}

This indicates that GitHub Actions is usable for our sequential MooBench execution. Since we do not have control over GitHubs measurement environment, we cannot generalize this for other benchmarks. Furthermore, since GitHubs investment in hosting might change, we do not know whether this low noise and good performance will be stable. Nevertheless, our results indicate that for benchmarks with low parallelization effort, GitHub Actions is likely to be a suitable measurement environment.

\subsection{Parallel Measurements}

To further examine the behavior of the different environments, we also compared the MooBench measurements for 2, 4, 8, and 12\footnote{16 Threads leads to cancellation of the GH actions process.} threads on GitHub Actions, Raspberry Pi, i7-4770, and Ryzen 5700G. Unfortunately, Kieker's queue overflows when using two threads in MooBench, since too many records are created and inserted into the queue. By default, Kiekers \lstinline'SPBlockingPutStrategy' inserts the records into an \lstinline'MpscArrayQueue' \cite{strubel2016refactoring}. This "only works correctly if at most one producer accesses the queue".\footnote{From Javadoc: \url{https://github.com/kieker-monitoring/kieker/blob/main/monitoring/core/src/kieker/monitoring/queue/putstrategy/SPBlockingPutStrategy.java}}

Therefore, we set the strategy to \lstinline'YieldPutStrategy' in MooBench. Since Kieker should be usable with parallel threads for third-party users, we made this the overall default configuration.

\begin{figure}
  \includegraphics[width=8cm]{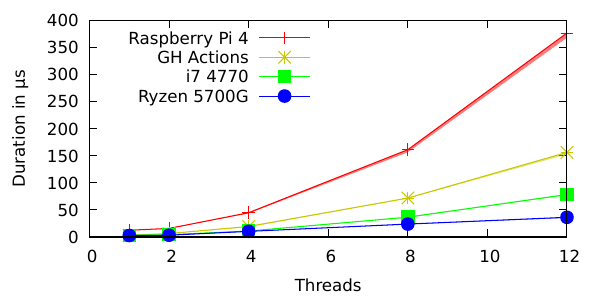}
  \caption{Evolution of Duration and Standard Deviation with Thread Count}
  \label{fig:threads}
\end{figure}

The average measurement values and standard deviations are depicted in Figure~\ref{fig:threads}. This shows that while GitHub Actions offers fast and stable measurements for sequential execution, parallelized execution slows down faster than on the i7-4770 environment. However, we still do not see a strong increase in standard deviation (in the graph, the standard deviation is even barely visible). Therefore, for the MooBench use case, GitHub Actions can be considered an appropriate runtime environment.

\section{Related Work}

Related work exists in two fields: Examination of the measurement stability in different execution environments and overhead comparisons of MooBench.

\subsection{Stability of Execution Environments}

Bulej et al. \cite{bulej2020duet} examined how performance changes can be detected using parallel benchmarking of two versions in the cloud. They found parallel execution of two versions can efficiently detect performance changes.
Laaber et al. \cite{laaber2019software} researched the variability in cloud measurements on AWS, GCE, and Azure. They found that variability varies across providers and that, depending on the benchmark and the cloud provider, changes can be detected using multiple runs.
In contrast to these works, we also aim to show the long-term evolution of our microbenchmark, instead of only detecting small changes.
% Our results show that MooBench on GitHub is able to detect those small changes.

Volpert et al. \cite{volpert2024empirical} researched how isolation capabilities of containers in cloud environments can reduce measurement noise. They found that CPU-bound workloads are isolated well by containers as workloads do not lead to full utilization of the CPU. For I/O-bound operations, they find that measurements are less stable. Since our measured deviations are comparably low, we assume that this is due to efficient isolation capabilities used by GitHub.

\subsection{Overhead Comparisons}

Analyzing past measurement results, that were executed on varying software environments, we see changes in the measured overhead. While Waller et al. \cite{waller2014application} measured an overhead of 60 $\mu s$ per method execution, Strubel and Wulf measured 3.27 $\mu s$ with the TCP writer (no binary writer values given) after their queue refactoring \cite{strubel2016refactoring}. Our prior measurements reported 3.4 $\mu s$ \cite{reichelt2021overhead}, $3.70\,\mu s$ (without warmup removal) \cite{reichelt2023more}, and 3.35 $\mu s$ \cite{reichelt2024overhead}.

To stabilize the measured overhead, Knoche and Eichelberger \cite{knoche2018using} proposed using the Raspberry Pi for reproducible overhead measurement. While they reported $121.5\,\mu s$ (Raspberry Pi 3), our later works reported $73.8\,\mu s$ \cite{reichelt2021overhead}, $17.9\,\mu s$ (without warmup removal) \cite{reichelt2023more} and $12.69\,\mu s$ \cite{reichelt2024overhead} overhead (all on Raspberry Pi 4). These tremendous changes in overhead measurement show that even if using the Raspberry Pi as platform for reproducability gives some stability, measurements change due to JVM and Kieker configuration changes.

\section{Summary}

In this paper, we examined how performance changes can be detected for different execution environments in MooBench. We found that while GitHub Actions is easy to set up, it has higher variation than bare metal machines.
In future work, we plan to evaluate the overhead of different instrumentation technologies, including the Java class file API.\footnote{\url{https://openjdk.org/jeps/457}}
% Auch https://stackoverflow.com/questions/67278973/how-to-trace-a-java-process-with-ebpf-bcc ?

\setlength\bibitemsep{3pt}
\printbibliography

\end{document}